\documentclass[12pt, a4paper, fleqn]{article}
\usepackage[utf8x]{inputenc}
\usepackage{amsmath}
\usepackage{amsfonts}
\usepackage{amssymb}
\usepackage{empheq}
\usepackage[sort&compress,numbers]{natbib}
\usepackage{doi}
\usepackage{esvect}
\usepackage{cancel}
\usepackage{braket}
\usepackage{hyperref}
\usepackage{xcolor}
\usepackage{tikz}

\usepackage[top=20mm, bottom=20mm, left=25mm, right=25mm]{geometry}

\newcommand{\p}{\partial}
\newcommand{\f}{\frac}
\newcommand{\s}{\sqrt}

\newcommand{\be}{\beta}

\newcommand{\D}{\Delta}

\newcommand{\bz}{\bar{z}}

\renewcommand{\c}[1]{\mathcal{#1}}

\newcommand*{\vol}[2]{\sqrt{#1}d^2{#2}\,\,}

\newcommand{\sph}{S^2}

\newcommand{\szego}{Szeg\"o\ }

\numberwithin{equation}{section}
	
\hypersetup{
	colorlinks   = true, %Colours links instead of ugly boxes
	urlcolor     = blue, %Colour for external hyperlinks
	linkcolor    = blue, %Colour of internal links
	citecolor   = red %Colour of citations
}

\newcommand{\wt}{\widetilde}

\newcommand{\lthree}{-0.28\times 10^{-3}}
\begin{document}
\

\begin{flushleft}
	{\bfseries\sffamily\Large 
		Liouville perturbation theory\\ for  Laughlin state and Coulomb gas
		\vspace{1.5cm}
		\\
		\hrule height .6mm
	}
	\vspace{1.5cm}
	
	{\bfseries\sffamily 
		Nikita Nemkov$^{1,2}$ and Semyon Klevtsov$^{3}$
	}
	\vspace{3mm}

	{\textit{\ \ 
			$^1$Mathematisches Institut, Universit{\"a}t zu K{\"o}ln, Weyertal 86-90, 50931 K{\"o}ln, Germany
			\\ \ \ 
			$^2$Institut f{\"u}r theoretische Physik, Universit{\"a}t zu K{\"o}ln, Zulpicher Str.\ 77a, 50937 K{\"o}ln, Germany\\ \ \
			$^3$IRMA, Universit\'e de Strasbourg, UMR 7501, 7 rue Ren\'e Descartes,
			67084 Strasbourg, France
		}}
		\vspace{2mm}
		
		{\textit{\ \ E-mail:} 
				\texttt{nemkov@mi.uni-koeln.de, 
				klevtsov@unistra.fr
			}}
		\end{flushleft}
		\vspace{7mm}
		
	{\noindent\textsc{Abstract:} We consider the generating functional (logarithm of the normalization factor) of the Laughlin state on a sphere, in the limit of a large number of particles $N$. The problem is reformulated in terms of a perturbative expansion of a 2d QFT, resembling the Liouville field theory. We develop an analog of the Liouville loop perturbation theory, which allows us to quantitatively study the generating functional for an arbitrary smooth metric and an inhomogeneous magnetic field beyond the leading orders in large $N$. 
	}

		\clearpage
		
		\hrule 
		{\hypersetup{linkcolor=black}
		\tableofcontents
		}
		\vspace{5mm}
		\hrule
		\vspace{5mm}

\section{Introduction}
In order to study the Laughlin wave function \cite{L} analytically it is customary to define it on a compact Riemann surface. The convenience of this approach, pioneered in \cite{H,HR,WN} is that one does not need to impose the boundary conditions, which may break holomorphy of the wave function in the bulk. Another advantage stems from the fact that Riemann surfaces are equipped with various discrete and continuous parameter spaces, such as the genus, moduli, Riemannian metric and the response to a change in these parameters allows one to extract analytically a wealth of physical information, such as the Hall conductance, Hall viscosity, bulk central charge, corrections to the static structure factors, etc \cite{ASZ,ASZ1,K,BR1,CLW,CLW1,CP,FK,FKer,FS,GHM,GS,H1,HS,H2,KW,KMMW,K16,LCW,L1,T2,RG,RR1999,R,RR,Son,TV2,ZMP}. Adiabatic geometric transport of the vector bundles of quantum Hall states over these parameter spaces gives rise to the Chern-Simons description of the quantum Hall effect, for recent developments see \cite{AG1,AG2,AG4,BR,KMMW}.

After the holomorphic part of the Laughlin state on a compact surface has been defined, the main problem is to compute its $L^2$ normalization, in the limit where the number of particles $N$ goes to infinity. This problem is closely related to the computation of the partition function of the Coulomb gas at integer values of inverse temperature $\beta$, at least in the case of the spherical topology.  

In this paper we study the normalization factor of the Laughlin state $Z[W]$, Eq.\ \eqref{Zdef} on the sphere as a functional of the background metric and inhomogeneous magnetic field, at large $N$. The logarithm of the  normalization factor can be used to compute the density-density correlation functions and for this reason it is also called the \emph{generating functional}. In the case of the integer quantum Hall state ($\beta=1$), the large $N$ expansion of the generating functional can be extracted rigorously \cite{K,KMMW} from the Bergman kernel expansion for high powers of the magnetic line bundle \cite{Z,Xu,MM}. In the fractional case $\beta>1$, the leading terms of the generating functional can be computed \cite{CLW,FK} using the $\beta$-ensemble Ward identity method of Zabrodin-Wiegmann \cite{ZW} or with the help of the free field construction of Moore-Read \cite{MR}. The rigorous proof of the large $N$ expansion of the generating functional is available up to the order $N$ due to \cite{LS}, see also \cite{BBNY}. For the rigorous approach to the Ward identities we refer the reader to \cite{AHM} and for the work on structure factors in the Coulomb gas we refer to \cite{KMST}.

The goal of this paper is twofold. On the one hand, we would like to systematically incorporate the case of inhomogeneous magnetic field $B$. This is an important issue since the corresponding contributions to the generating functional are not small (of order $B\log B$). We derive the corresponding terms up to the order $N^0$ and compare them to the result conjectured in \cite[Eq.\ 130]{CLW1}. On the other hand both in the Ward identity and free field methods it is hard to control the subleading corrections of the large $N$ expansion. In Ref.\
 \cite{FK} the terms up to the order $N^0$ of the generating functional were derived, and remainder term for the large $N$ expansion was written as a path intergral in an interacting 2d QFT, but only general arguments were given as to why this term is expected to be small, and local, at large $N$. In this paper we develop a perturbative scheme inspired by the semiclassical analysis of the Liouville theory \cite{Zamolodchikov:1995aa,Zamolodchikov:2001ah} in order to compute the order $1/N$ contribution to the generating functional. 

Let us now describe our main result. From general principles we expect the following formula for the large $N$ expansion of the generating functional up to the order $1/N$,

\hspace*{-1.5cm}\vbox{
\begin{multline}\log Z=-\f1{2\pi}\sigma_H\iint_{(\sph)^2} B\D_g^{-1}B-\f1{2\pi}2\varsigma_H\iint_{(\sph)^2} B\D_g^{-1}R+\frac1{2\pi}\cdot\f{c_H}{48}\iint_{(\sph)^2} R\D_g^{-1}R\\
-\f{1}{2\pi}\int_{\sph} \s{g}d^2z\left[\frac{2-\beta}{2\beta} B\log B+\left(\f1{24}-\frac{(\beta-2)(\beta-2s)}{8\beta}\right)R\log B+\frac c{48}(\log B)\D_g\log B\right]\\-\f{1}{2\pi}\int_{\sph}\vol{g}{z}\big(c_1(\beta)(\D_g\log B)^2B^{-1}+c_1'(\beta)R(\D_g \log B)B^{-1}+c_1''(\beta)R^2B^{-1}\big)+O(N^{-2})
\label{main}
\end{multline}
}

\noindent First let us comment on the order of terms in this expansion. The magnetic field is assumed to be large $B\sim N$ while the scalar curvature $R$ of the background metric is of order $N^0$. In this sense the first term here is of order $N^2$, the terms in the last two lines vanish when $N\to\infty$, while the other terms have the orders $N\log N, N,\log N$ and $N^0$. 

The first line in \eqref{main} contains so-called anomalous terms, which were computed in \cite{K,CLW,FK}. The coefficients of the anomalous terms are
\begin{align}
\sigma_H=1/\beta,\qquad \varsigma_H=\frac14-\f{s}{2\beta},\qquad c_H=1-\f{3(\beta-2s)^2}{\be}, \label{cHall}
\end{align}
respectively the  Hall conductance, the Hall viscosity and the (bulk) Hall central charge. Parameter $s$ is the so-called gravitational spin, introduced according to \eqref{Wdef}. Since in this paper we will not be concerned with the anomalous terms, we refer to \cite{FK,FKZ3} for the proper definitions of the double integrals and regularized Green functions $\Delta_g^{-1}$ in Eq.\ \eqref{main}. The second line involves the terms which are single integrals of  functions of the scalar curvature, the magnetic field and their derivatives. Interestingly, the value of the parameter $c$ here coincides with the value of the bulk central charge $c_H$ at $s=1$
\begin{align}
c=1-3\f{(\be-2)^2}{\be} \ . \label{clabel}
\end{align}
This value was also conjectured to be relevant to Liouville theory in \cite{Vafa}. 

Finally, the last  line in \eqref{main} contain the terms which are of order $N^{-1}$ and smaller. We expect these terms to be given by local integrals of the densities depending on $B$, $R$ and their derivatives, with the expansion being essentially the derivative expansion. The form of the terms in the last line is constrained by the diffeomorphism invariance of the generating functional. The coefficients $c_1(\be),c_1'(\be),c_1''(\be)$ are real-valued functions of $\beta$ whose value is only known at $\beta=1$, for constant magnetic field \cite{K}. In this paper we use our Liouville perturbation theory in order to compute these coefficients as the Laurent series around $\beta=0$. We establish that their principal part is a simple pole in $\beta$ and compute several terms in the small-$\beta$ expansion perturbatively, see Eqns.\ (\ref{c1exp},\ref{c1exp'},\ref{c1exp''}). We also argue that the coefficients in the higher order terms of the large $N$ expansion are also given by the Laurent series with simple poles at $\beta=0$. 

While this is not our main goal, our method can also be used to compute the generating function in the regime $N\to\infty$ and $\beta\to0$ with $\beta< C N^{-\epsilon}$, where $0<\epsilon\leqslant1$. The case of $\epsilon=1$ corresponds to the generalized Gibbs ensemble in the probabilistic approach to K\"ahler-Einstein metrics \cite{Ber}. 

The paper is organized as follows. In sec.\ \ref{sec gen} we define the generating functional and introduce the notations relevant for Eq.\ \eqref{main}. In sec.\ \ref{sec qft} we recall the relation between the Laughlin state on curved backgrounds and free fields. We then define an interacting QFT that can be used to compute the generating functional. In sec.\ \ref{sec leading} we make our proposal more precise by emphasizing similarity of our model with the Liouville field theory. We find a convenient reparametrization of the path integral that naturally separates the leading order terms of the generating functional from the subleading ones. In sec.\ \ref{sec loops} we develop a perturbation theory which in principle allows to compute the subleading terms as the Laurent expansions in $\be$. We focus on the first non-trivial coefficient and compute it up to three loop orders. In sec.\ \ref{sec disc} we review our results and outline directions for future work.
\section{Generating functional \label{sec gen}}
\subsection{Definition}
The main object in the present paper is the generating functional defined as the logarithm 
\begin{align}
F[W]=\log Z[W] \label{Fdef}
\end{align}
of the $N-$fold integral over a sphere with respect to the volume form $dv(z)=\sqrt gd^2z$
\begin{align}
Z[W]=\int_{(\sph)^N} \prod_i^N dv(z_i) \left|\psi(z_1,...,z_N)\right|^2 \label{Zdef}
\end{align}
of the mod squared of the Laughlin wave function $\psi$ at filling fraction $\be^{-1}$ with $\beta\in\mathbb Z_+$,
\begin{align}
\psi(z_1,..,z_N)=\prod_{i<j}^N (z_i-z_j)^{\beta}\prod_i^N e^{\f12W(z_i,\bar{z}_i)} \ . \label{Psidef}
\end{align}
Here the locally-defined potential function $W(z,\bar z)$  
\begin{align}
W(z,\bar z)=K(z,\bar z)-s\log\s{g}(z,\bar z) \label{Wdef} 
\end{align}
is the sum of the potential function $K$ for the magnetic field $B=-\f12\D_g K$, which is strictly positive $B>0$, and the logarithm of the metric where the coefficient $s\in\frac\be2\mathbb Z$ is the so-called gravitational spin of the Laughlin state. For the fully filled Laughlin ground state the quantized flux of magnetic field is related to the number of particles as follows
\begin{align}
N_\phi=\frac1{2\pi}\int_{\sph} B\, \s{g}d^2z = \be (N-1)+2s \ .\label{Bflux}
\end{align}
The generating functional encodes the response of the ground state to external electromagnetic and gravitational fields and contains a wealth of information about the state. For instance, in the spherical geometry one application of the generating functional is to compute the density-density correlation functions \cite{ZW}.

\subsection{Structure of the large $N$ expansion}
Here we start with a reminder of what is known about the large $N$ expansion of the generating functional and introduce several conventions and notations along the way. First, we note that one can think of $F[W]$ as a functional of $W$, or more precisely of $W+\log\sqrt g$, as is apparent from the definition. On the other hand it is useful to think of it as a functional of the magnetic potential $K$ and the metric $g$, as well as their derivatives -- the magnetic field $B$,  scalar curvature $R$ etc. The generating functional naturally splits (see e.g. \cite{K16} for a review) into the anomalous $\c{F}_A$ and exact parts $\c{F}_E$
\begin{align}
F[W]=\c{F}_A[g, B]+\c{F}_E[g, B] \ . \label{FAE}
\end{align}
The anomalous part can be formally written as a non-local expression
\begin{multline}
\c{F}_A[g, B]=-\f1{2\pi}\sigma_H\iint_{(\sph)^2} B\D_g^{-1}B-\f1{2\pi}2\varsigma_H\iint_{(\sph)^2} B\D_g^{-1}R+\f{c_H}{96\pi}\iint_{(\sph)^2} R\D_g^{-1}R \ . \label{cFAdef}
\end{multline}
This is a schematic formula, for details of the definition 	we refer to \cite{FK,K16}. This expression is a combination of gauge, gravitational and mixed gauge-gravitational anomaly functionals and the corresponding coefficients, depending on $\beta$ and $s$ encode the Hall conductance $\sigma_H=1/\beta$, the Hall central charge $c_H=1-\f{3(\be-2s)^2}{\be}$ and the third coefficient $\varsigma_H=\frac14-\f{s}{2\be}$, related to Hall viscosity. 

According to 
\eqref{Bflux} the magnetic field $B$ is of order $N$ for $N$ large, while the curvature $R$ is of order one, hence anomalous part contains the terms of order $ N^2,N^1$ and $N^0$ of the large $N$ expansion. In turn, the exact part $\c{F}_E$ is a local (single-integral) gradient expansion in terms of the scalar curvature and magnetic field with the leading terms given by
\begin{multline}
\mathcal{F}_E[g,B]=-\f{1}{2\pi}\int_{\sph} \s{g}d^2z\left[\left(\f1{\be}-\f12\right) B\log B+\left(\left(\f1\be-\f12\right)\f{\be-2s}{4}+\f1{24}\right)R\log B-\right.\\\left.\left(\f1{4\be}-\f{13}{48}+\f{\be}{16}\right)(\log B)\D_g\log B\right]+\mathcal R_N, \label{cFexactLeading}
\end{multline}
where $\mathcal R_N$ is of order $O(N^{-1})$
This equation was proven rigorously at $\be=1$ in \cite{KMMW} (and \cite{K} for constant magnetic field) and conjectured for any $\be$ in \cite[Eq.\ 130]{CLW}. The difficult part is to control the remainder terms $\c{R}_N$ well enough to prove that they are subleading. We note that both formulas \eqref{cFAdef} and \eqref{cFexactLeading} are written as functionals that depend separately on the metric, magnetic field and spin. However their sum can in fact be recombined into a functional of $W+\log\sqrt g$ alone. 

A word about conventions. As exemplified by equation \eqref{FAE} throughout the paper we use calligraphic symbols like $\c{F}_E$ for functionals that depend on magnetic field and metric separately and reserve plain symbols like $F[W]$ for true functionals of potential $W$. Note also a complementary notation for the functionals of $W$ discussed around \eqref{FWFgH}.
In sec.\eqref{sec qft} we propose a path integral formulation of the generating functional, which produces the leading order terms of the large $N$ expansion in a way that unifies \eqref{cFAdef} and \eqref{cFexactLeading}. 

\subsection{Notation}
\subsubsection*{Differential geometry}
We will assume that the metric in conformal coordinates is $ds^2=2g_{z\bar z}dzd\bar z$ and introduce the scalar Laplacian and the scalar curvature
\begin{align}
\D_g=\f{4}{\s{g}}\p_z\p_{\bar z},\qquad R=-\D_g\log \s{g} \ ,
\end{align}
with $\s{g}=2g_{z\bar z}$. Conventions are such that $\int_{\sph} \vol{g}{z}R=4\pi\chi$ with $\chi$ being the Euler characteristic ($\chi=2$ for a sphere).

We will frequently consider a pair of metrics related by a Weyl rescaling with some conformal factor $\Omega(z,\bz)>0$. In this case we will write $g_\Omega$ for the rescaled metric with
\begin{align}
\left(g_{\Omega}\right)_{z\bar z}=\Omega g_{z\bar z},\qquad \s{g_\Omega}=\Omega\s{g} \label{gOmegadef}
\end{align}
and also
\begin{align}
\D_\Omega=\Omega^{-1}\D_g,\qquad R_\Omega=\Omega^{-1}\left(R-\D_g\log\Omega\right) \ .
\end{align}
We normalize delta-functions on a curved sphere with respect to the volume element
\begin{align}
\int_{\sph} \vol{g}{z} \delta(z,w)=1 .
\end{align}
\subsubsection*{Effective magnetic field}
We will refer to the following combination of magnetic field and scalar curvature \begin{align}
H=B+\f{1-s}{2}R =-\frac12\Delta_g (W+\log \sqrt{g})\ . \label{Hdef}
\end{align}
as the {\it effective magnetic field}. The flux of $H$ is
\begin{align}
N_H=\frac1{2\pi}\int_{\sph} \s{g}d^2z\,\, H = \be (N-1)+2 \ .\label{Hflux}
\end{align}
Since $H>0$, we can introduce a fictitious metric $g_H$ as is suggested by \eqref{gOmegadef}
\begin{align}
\s{g_H}=H\s{g} \label{gHdef}
\end{align}
and its normalized version
\begin{align}
\s{g_h}=h\s{g},\quad h=\frac{H}{N_H}\ , \label{ghdef}
\end{align}
which has a fixed area in the limit $N\to\infty$
\begin{align}
\f{1}{2\pi}\int_{\sph}\vol{g}{z} h=\f{1}{2\pi}\int_{\sph}\vol{g_h}{z}=1 \ .
\end{align}
Note that since $\s{g}\D_g$ is in fact independent of the metric the combination $\s{g}H$ is a function of $W+\log \sqrt{g}$ alone. Later we will mostly think of the generating functional as the functional of $g_H=N_H g_h$ and with a slight abuse of notation write
\begin{align}
F[W]=F[g_H] \ . \label{FWFgH}
\end{align}

\section{Laughlin state and 2d QFT \label{sec qft}}  

\subsection{Laughlin state from free QFT}
Consider a free scalar field theory with action
\begin{align}
\c{S}_0[\phi, g, B]=\f{1}{4\pi}\int_{\sph} \vol{g}{z}\left[-\phi\Delta_g\phi+i\s{\f{8}{\be}}\left(B+\f{\be-2s}{4}R\right)\phi\right] \ . \label{S0}
\end{align}
Laughlin's wavefunction \eqref{Psidef} can be interpreted as a correlation function of the electron operators \cite{MR},
\begin{align}
V_e(z,\bar z)=e^{i\s{2\be}\phi(z,\bar{z})} \label{Vedef}
\end{align}
in this theory. Precisely \cite{FK},
\begin{align}
\left|\psi(z_1,...,z_N)\right|^2=e^{\c{F}_A[g,B]}\Braket{\prod_iV_e(z_i, \bar z_i)} =e^{\c{F}_A[g,B]}\int \c{D}_g\phi\,\, e^{-\c{S}_0[\phi,g,B]}\prod_i e^{i\s{2\be}\phi (z_i,\bar z_i)} \ . \label{psiCorr}
\end{align}
Note that the correlation function here is not normalized. 

The scalar field is assumed to be compact $\phi\equiv \phi+2\pi r_c$ with the radius of the circle $r_c=\s{\be/2}$. The electron operators are invariant under shifts of $\phi$ by multiples of $2\pi r_c$. As can be seen by separating the zero-mode integration in \eqref{psiCorr} this correlation function is only well-defined (invariant under $\phi\to\phi+2\pi r_c$) and non-vanishing if the magnetic field has flux \eqref{Bflux}.  Conversely, given a magnetic field with flux \eqref{Bflux} only the correlation function with $N$ electron operators is non-zero.

Substituting \eqref{psiCorr} in \eqref{Fdef} and canceling the common anomaly factors one finds 
\begin{align}
e^{\c{F}_E[g,B]}=\int_{(\sph)^N} \prod_i \vol{g}{z_i} \Braket{\prod_iV_e(z_i, \bar z_i)}= \Braket{\left(\int_{\sph} \vol{g}{z} \prod V_e(z, \bar z)\right)^N} \label{FEascorr} \ .
\end{align}

The holomorphic part of Laughlin's wavefunction can be identified with conformal block of a free boson CFT. We emphasize that theory \eqref{S0} is strictly speaking not a CFT as the presence of the magnetic field breaks the conformal invariance. This distinction might not seem significant at the level of free fields but it becomes essential in the interacting theory that we will now introduce.
\subsection{Generating functional from interacting QFT}
Let us define a modified version of the free action \eqref{S0}
\begin{align}
\c{S}[\phi, g, B]=\f{1}{4\pi}\int_{\sph} \vol{g}{z}\left(-\phi\Delta_g\phi+i\s{\f{8}{\be}}\left(B+\f{\be-2s}{4}R\right)\phi+4\pi \mu e^{i\s{2\be}\phi}\right)\ , \label{SB}
\end{align}
which can be thought of as a perturbation by the electron operator
\begin{align}
\c{S}=\c{S}_0+\mu \int_{\sph} \vol{g}{z}V_e(z,\bar z) \ .
\end{align}
Now consider the partition function of this theory and formally expand it in powers of $\mu$
\begin{align}
\int \c{D}_g\phi\,\, e^{-\c{S}[\phi,g,B]}=\sum_{n=0}^{\infty}\f{(-\mu)^n}{n!}\int \c{D}_g\phi\,\, e^{-\c{S}_0[\phi,g,B]}\left[\int_{\sph} \vol{g}{z} V_e(z,\bar z)\right]^n \ .
\end{align}
The right hand side here is a sum of the integrated correlators in the free theory. In fact only one term with $n=N$ satisfies the charge neutrality and is non-vanishing. Comparing to \eqref{FEascorr} we find that the partition function of interacting theory \eqref{SB} with properly adjusted flux of the magnetic field \eqref{Bflux} computes the generating functional of the Laughlin state
\begin{align}
e^{\c{F}_E[g,B]}=\int \c{D}_g\phi\,\, e^{-\c{S}[\phi,g,B]} \ . \label{mainconj}
\end{align}
This observation is the starting point for most of our further developments. Note that we have absorbed an inessential constant $(-\mu)^N/N!$ into $\c{F}_E$ and we will keep this convention in the following.

\section{Leading terms from Liouville field theory \label{sec leading}}

\subsection{Relation to LFT}
Any non-trivial QFT is only tentatively defined by its classical action. To shape our proposal we will draw on the similarity between \eqref{SB} and the Liouville field theory (LFT) defined by action
\begin{align}
S_L[\phi, g]=\f{1}{4\pi}\int_{\sph} \vol{g}{z}\left(-\phi\Delta_g\phi+QR\phi+4\pi \mu e^{2b\phi}\right) \ , \label{SL}
\end{align}
where $Q=b+b^{-1}$. In fact one finds
\begin{align}
\c{S}[\phi, g, B]=S_L[\phi, g]-\f{1}{2\pi b}\int_{\sph} \vol{g}{z}H\phi \label{StoSL}
\end{align}
if we identify
\begin{align}
b=i\s{\f{\be}{2}} \label{bdef}\ ,
\end{align}
and we will comment on the choice of imaginary $b$ below.
To facilitate comparison with LFT in the current section we always use $b$ instead of $\be$. We will return to the conventional FQHE parameter $\be$ in sec.\eqref{sec loops}.

LFT is known to be a conformal field theory with central charge 
\begin{align}
c=1+6Q^2 \ . \label{cQ}
\end{align}
In particular this implies that under a Weyl rescaling of the metric $g\to g_\Omega$ its partition function
\begin{align}
Z_L[g]=\int \c{D}_g\phi\,\,e^{-S_L[\phi,g,B]}
\end{align}
transforms as
\begin{align}
Z_L[g_\Omega]=Z_L[g]\,\,e^{c\,\,A[g,\Omega]} \ . \label{ZLtransform}
\end{align}
Here $A[g,\Omega]$ is the Polyakov Liouville functional
\begin{align}
A[g,\Omega]=-\f{1}{96\pi}\int_{\sph} \s{g}d^2z\Big(\log \Omega\,\,\D_g \log \Omega-2R\log \Omega\Big) \ .
\end{align}
To obtain \eqref{ZLtransform} one uses the transformation properties of the path integral measure and the action under a Weyl rescaling
\begin{align}
\f{\c{D}_{g_{\Omega}}\phi}{\c{D}_{g}\phi}=e^{A[g,\Omega]}, \qquad S_L[\phi, g_\Omega]=S_L[\phi+\f{Q}{2}\log\Omega,g]-6Q^2 A[g,\Omega] \ . \label{SLtransform}
\end{align}
To derive the transformation law for the action one needs the following relation
\begin{align}
\left[ e^{2b\phi}\right]_{g_\Omega}=e^{b^2\log\Omega}\left[ e^{2b\phi}\right]_{g} \ , \label{etransform}
\end{align}
which is a part of the quantization prescription of the classical theory \eqref{SL}. The interaction exponent is assumed to be normal-ordered implying that the 	short-distance singularities are regulated with a reference to a length scale introduced by metric $g$. When the metric is Weyl rescaled the interaction term transforms non-trivially \eqref{etransform}.

We now emphasize several distinctions between LFT and our model \eqref{StoSL}. Note that in the last step of deriving \eqref{ZLtransform} from \eqref{SLtransform} one needs to shift the path integral variable $\phi\to\phi-\f{Q}{2}\log\Omega$. This is possible literally when $Q\in\mathbb{R}$, i.e. $c>25$ and $\phi$ is not compact. This is the best studied regime of LFT. In contrast, the case that is directly relevant for our discussion \eqref{bdef} is $c\le1$ with $Q^2\in\mathbb{Q}$. By definition LFT is still conformal and obeys \eqref{ZLtransform} but its description in terms of the path integral can be more subtle \cite{Harlow:2011ny}. However, precisely when $Q^2$ is rational the Liouville theory is believed to be essentially similar to generic $\operatorname{Re} c>1$ and in particular obtainable by the analytic continuation from that region \cite{Ribault:2015sxa}. This suggests that the relevant path integral allows for complex deformations of the integration contour. We will assume this to hold also for the path integral with the magnetic field \eqref{StoSL}. An additional subtlety is that in \eqref{StoSL} the free boson is compact and so the zero mode integration is over a finite range. This seems to prevent us from making constant shifts in $\phi$. However, the result of such a shift on the generating functional \eqref{mainconj} is to rescale the parameter $\mu$ which simply leads to a constant shift in $\c{F}_E$ and is not important. As we will see shortly, the assumptions we have made are justified a posteriori as they allow to correctly reproduce the leading order terms of the generating functional.

Finally, we stress that the true power of LFT is grounded in conformal symmetry. In this sense although the modification by the additional background term in \eqref{StoSL} might seem mild it actually breaks the conformal invariance and is therefore essential. Nevertheless, the analogy with LFT will be a very useful reference point for our further developments. 

\subsection{Generating functional: anomalies and leading terms}

As the first application of our proposal we will derive the leading terms in the large $N$ expansion of the generating functional. We begin with the path-integral representation \eqref{mainconj} for the exact part of the generating functional $\c{F}_E$ and make the Weyl rescaling from metric $g$ to $g_H$. As we will see this transformation naturally decouples the leading terms in the generating functional from $O(N^{-1})$ tail. Using relation to the LFT action \eqref{StoSL} and the transformation property of the latter \eqref{SLtransform} it follows
\begin{align}
\c{S}[\phi-\f{Q}{2}\log H,g,B]=S[\phi,g_H ]+6Q^2A[g,H] \ ,
\end{align}
where we introduced a new action $S$ that only depends on $g_H$,
\begin{align}
S[\phi, g_H]=S_L[\phi,g_H]-\f{N_H}{2\pi b}\int_{\sph} \vol{g_h}{z} \phi \ . \label{SgH}
\end{align}
Shifting the path integral variable $\phi\to\phi-\f{Q}{2}\log H$, accounting for the measure anomaly \eqref{SLtransform} and an additional contribution from the second term in \eqref{SgH} one finds
\begin{align}
\c{F}_E[g,B]=\wt{\c{F}}_E[g,B]+F_E[g_H] \label{FBFH}
\end{align}
with 
\begin{multline}
\wt{\c{F}}_E[g,B]=-c A[g,H]+\f{Q}{4\pi b}\int_{\sph} \vol{g}{z}H\log H =\\\f{Q}{4\pi b}\int_{\sph} \vol{g}{z}H\log H+\f{c}{96\pi}\int_{\sph} \vol{g}{z}\Big(\log H\left(\D_g \log H\right)-2R\log H\Big) . \label{cFtE}
\end{multline}
Here $c$ is the central charge \eqref{cQ} and $F_E[g_H]$ is the generating functional associated with action \eqref{SgH} (explicitly defined later in \eqref{Ztoloop}).

Passing from the original metric $g$ to metric $g_H$ in the path integral which leads to decomposition \eqref{FBFH} is very useful. First, using definition of $H$ \eqref{Hdef} one can check that $\wt{\c{F}}_E$ reproduces the leading exact terms of the generating functional expansion \eqref{cFexactLeading}
\begin{align}
\wt{\c{F}}_E[g,B]=\c{F}_E[g,B]+O(N^{-1}) \ .
\end{align}
The remaining part of the exact term $F_E[g_H]$ in \eqref{FBFH} is then expected to be capture the remainder terms $\c{R}_N$ in \eqref{cFexactLeading} and be subleading. We will argue that this is indeed the case in the next section using path integral representation for $F_E[g_H]$.

Here we would also like to note that $\wt{\c{F}}_E$ can be neatly combined with the anomaly terms \eqref{cFAdef} into a function of $g_H$ alone
\begin{multline}
F_A[g_H]=\c{F}_A[g, B]+\wt{\c{F}}_E[g, B]=\\\f{N_H^2}{4\pi b^2}\iint_{\sph} \D_h^{-1}-\f{N_HQ}{4\pi b}\iint_{\sph} \D_h^{-1}R_h+\f{c}{96\pi}\iint_{\sph} R_h\D_h^{-1}R_h \ . \label{FAH}
\end{multline}
To put it differently, this expression reproduces both the standard from of the anomaly terms \eqref{cFAdef} and the leading orders of the exact part \eqref{cFexactLeading}. As a bonus this formula also nicely interprets coefficients in \eqref{cFexactLeading} in terms of Liouville central charge $c$. We used this observation in \eqref{main}.
\section{Subleading corrections \label{sec loops}}
\subsection{Setup and conventions}
In the previous section we have found that there is a natural split of the generating functional 
\begin{align}
F[g_H]=F_A[g_H]+F_E[g_H] \ ,
\end{align}
where $F_A[g_H]$ defined by \eqref{FAH} accounts for all leading terms of the large $N$ expansion (both anomalous and exact). The rest $F_E[g_H]$ is then expected to capture the remainder terms $\c{R}_N$ in \eqref{cFexactLeading} and vanish in the large $N$ limit. It is the purpose of the current section to explain why $F_E[g_H]$ is indeed subleading with respect to $F_A[g_H]$ and how it could in principle be computed in our approach.

We begin with the path integral representation of $F_E[g_H]$
\begin{multline}
e^{F_E[g, N]}=\int\c{D}_g\phi\,\,e^{-S[\phi,g, N]}=\\\int\c{D}_g\phi \,\,\exp\left[-\f{1}{4\pi}\int_{\sph} \vol{g}{z} \left(-\phi\D_g\phi+i\s{\f{2}{\be}}M\phi+4\pi \mu e^{i\s{2\be}\phi}\right)\right] \label{Ztoloop}\ .
\end{multline}
where for later convenience we introduced
\begin{align}
M=2N_H+\f{\be-2}{2}R \label{Mdef} \ .
\end{align}
Let us clarify the notations adopted. Firstly, we have switched back to $\be$ from the Liouville $b$ via \eqref{bdef}. More importantly, for brevity in this section we simply write $g$ instead of $g_h$ \eqref{ghdef}. So the dependence on the fictitious metric $g\equiv g_h$ in \eqref{Ztoloop} in fact describes the dependence on both the original geometry and the magnetic field. We also choose to explicitly reflect dependence on $N$ of the generating functional and the corresponding action in notations $F_E[g,N]$ and $S[\phi,g,N]$. Up to (inessential) renormalization of $\mu$ they are the same as in the previous section $F_E[g_h,N]=F_E[g_H]$ and $S[\phi,g_h,N]=S[\phi,g_H]$. 
\subsection{Gradient expansion}
The large $N$ expansion of the exact part of the generating functional is believed to be a local gradient expansion in $R$ and derivatives and its general form is  fixed by the diff-invariance,
\begin{multline}
F_E[g,N]=\f{c_{-1}(\be)N_H}{2\pi}\int_{\sph} \vol{g}{z}+\f{c_{0}(\be)}{2\pi}\int_{\sph} \vol{g}{z} R+\f{c_{1}(\be)}{2\pi N_H}\int_{\sph} \vol{g}{z}\,R^2+\\\f1{2\pi N_H^2}\int_{\sph} \vol{g}{z} \left(c_2(\be)R^3+c_2'(\be)R\D_g R\right)+O(N_H^{-2}) \ . \label{Fexp}
\end{multline}
Note that the integrals weighted by $c_{-1}(\be)$ and $c_0(\be)$ describe the total area and the Euler characteristic of the surface respectively. They thus do not exhibit any interesting dependence on the metric and we will simply omit them in what follows. The first non-trivial term in \eqref{Fexp} is thus
\begin{align}
F_E[g,N]\simeq \f{c_1(\be)}{2\pi N_H}\int_{\sph} \vol{g}{z} R^2+O(N_H^{-2}) \label{Fsimeq}
\end{align}
with $\simeq$ implying that the two trivial leading terms are omitted.

Computing coefficients $c_i(\be)$ seems to be a rather non-trivial task. Presently they are only known at $\be=1$ when the generating functional admits a determinantal representation and can be related to the Bergman kernel. Several orders in expansion of the full generating functional at $\be=1,s=0$ in homogeneous magnetic field were found in \cite[Eq.\ 6.6]{K}. Here we quote the result at $\beta=1$\footnote{Note that normalization of the scalar curvature used in \cite{K} is related to ours by a factor $1/2$.}
\begin{align}
F[g,N]\simeq-\f{5}{384\pi N}\int_{\sph} \vol{g}{z} R^2+O(N^{-2}) \ . 
\end{align}
To deduce the value of $c_1(\be=1)$ one must subtract the contribution of \eqref{cFtE} at this order. Substituting $H=N-1+\f{R}{2}$ which follows from \eqref{Hdef} at $\be=1,s=0$ gives
\begin{align}
\wt{\c{F}}_E[g,B]\simeq -\f1{96\pi N} \int_{\sph} \vol{g}{z} R^2 + O(N^{-2}) 
\end{align}
and hence
\begin{align}
c_1(\be=1)=- \f{1}{192} , \label{c1}
\end{align}
in our conventions here.

\subsection{Semiclassical expansion}
Given that $N_H$ is a large parameter in the action \eqref{Ztoloop} it may be tempting to treat the path integral semiclassically. However, this large parameter effectively only couples to the zero mode and hence does not suppress quantum fluctuations. In the LFT the semiclassical regime instead is $\be\to0$. Our strategy is to compute the gradient expansion to the order $1/N$ from the LFT perturbation theory, compare it to the expected form \eqref{Fexp} and extract the information about the coefficients $c_n(\beta)$ in perturbation theory around $\beta=0$. 

First, let us introduce a rescaled scalar field $\varphi=\s{2\be}\phi$ and rewrite \eqref{Ztoloop} accordingly
\begin{multline}
e^{F_E[g, N]}=\int\c{D}_g\varphi \,\,\exp\left[-\f{1}{8\pi \be}\int_{\sph} \vol{g}{z} \left(-\varphi\D_g\varphi+2iM\varphi+8\pi \mu \be e^{i\varphi}\right)\right] \label{Ztoloopvar}\ .
\end{multline}
This representation suggests to consider small $\beta$ limit at fixed $M$. Note that for the constant scalar curvature function $M$ \eqref{Mdef} is also a constant and therefore defined by its flux. As follows from \eqref{Hflux}
\begin{align}
\f{1}{2\pi}\int_{\sph} \vol{g}{z} M=2\be N \label{Mflux} \ .
\end{align}

Thus we are naturally led to consider the perturbative expansion in small $\beta$ and then large $M$, the latter turns out to correspond to the gradient expansion.  While this limit can be interesting on its own, our goal here is purely practical since this is the regime where perturbative calculations are possible. From the saddle-point arguments it follows that the coefficients $c_n(\be)$ are given by Laurent expansions in $\be$ with the principle part being by a simple pole,
\begin{align}
c_n(\be)=\sum_{k=-1}^\infty c_{n,k}\,\,\be^k \ . \label{cn}
\end{align}

In what follows we will focus on computing coefficient $c_1(\be)$ of the first non-trivial term in \eqref{Fexp}. Up to three loop orders we obtain a result of the form
\begin{align}
c_1(\be)=l_0(\be)+l_1(\be)+\be l_2+\be^2l_3+O(\be^3) \ , \label{c1init}
\end{align}
where $l_0(\be)$ comes from the saddle point action, $l_1(\be)$ from the one-loop determinant, and $l_2,l_3$ from the two- and three-loop diagrams. We note in advance that in our perturbation theory the saddle point and the one loop contributions mix together several orders of $\be$ while higher loops contribute to $c_1(\be)$ monomials with a power of $\be$ determined by the number of loops. There is no mixing of orders if $M$ is taken as a large expansion parameter instead of $N_H$ but in order to extract $c_1(\be)$ one needs to substitute \eqref{Mdef} and re-expand in $N_H$, introducing the mixing.

Since we are only interested in the $R^2$ term which does not involve derivatives of the curvature we can assume that the underlying geometry is that of a round sphere, and we can drop the derivatives of $R$ in our calculations.  

\subsection{Saddle point}
The classical equation of motion for the action in \eqref{Ztoloopvar} is
\begin{align}
4\pi\mu \be e^{i\varphi_s}=-M+\D_g\varphi_s \ . \label{EOM}
\end{align}
Recall that we are interested in the case of constant $M$ which allows for a constant solution
\begin{align}
\phi_s=-i\log\f{-M}{4\pi\mu \be} \ .
\end{align}
For $\phi_s$ to be real the argument of the logarithm must be a pure phase. This can always be achieved by an appropriate choice of $\mu$ which we assume. The action evaluated on the classical solution reads
\begin{align}
S[\phi_s,g,N]=\f1{8\pi \be}\int_{\sph} \vol{g}{z} \left[2M\log\f{-M}{4\pi\mu\be}+2M\right] \ .
\end{align}
To extract the contribution of the saddle point to $c_1(\be)$ we substitute \eqref{Mdef} and expand in $N_H$ 
\begin{align}
S[\phi_s,g,N]=const-N\log 4\pi\mu\be -\f{l_0(\be)}{2\pi N_H}\int_{\sph} \vol{g}{z}R^2+O(N_H^{-2}) \ , \label{Ss}
\end{align}
with
\begin{align}
l_0(\be)=-\f{(\be-2)^2}{32\be} \ .
\end{align}
We have kept dependence on $\mu$ in the classical action to emphasize that $e^{F_E[g]}\propto e^{-S[\phi_s]}\propto \mu^N$ consistently with \eqref{mainconj}. 
\subsection{Quantum fluctuations}
The fluctuations about the classical solution $\chi=\varphi-\varphi_s$ are described by the following action 
\begin{align}
S[\chi]=\f{1}{8\pi \be}\int_{\sph} \vol{g}{z}\Big[ -\chi\D_g\chi -2M\left(e^{i\chi}-i\chi-1\right)\Big] \ . \label{Schi}
\end{align}
We will use a standard perturbation theory expanding \eqref{Schi} around the the minimum of the potential and treating higher order terms perturbatively 
\begin{align}
S[\chi]=S_2[\chi]+\sum_{n\ge3}\f{g_n}{n!}\int_{\sph} \vol{g}{z} \chi^n,\qquad g_n= M \f{(i)^{n+2}}{4\pi\be} \ ,
\end{align}
where the quadratic part is
\begin{align}
S_2[\chi]=\f{1}{8\pi\be}\int_{\sph} \vol{g}{z}\left[ -\chi\D_g\chi +{M}\chi^2\right] \ . \label{S2}
\end{align}
Let us introduce the propagator of this Gaussian field by
\begin{align}
G_{M}(z,w)=\f{1}{2\be}\Braket{\chi(z)\chi(w)} \ .
\end{align}
The normalization is chosen so that $G_M(z,w)$ is $\beta$-independent. The propagator solves equation
\begin{align}
(-\D_g+{M})G_{{M}}(z,w)=2\pi \delta(z,w) \ . \label{GMeq}
\end{align}
We will need the expansion of the propagator near the diagonal which can be obtained for example from the off-diagonal expansion of the heat kernel (see for instance \cite{Groh:2011dw})
\begin{align}
G_{M}(z,w)=A_0 K_0\left(d\s{{M}}\right)+\f{A_1 d}{2\s{{M}}} K_1\left(d\s{{M}}\right)+\f{A_2 d^2}{4{M}}K_2\left(d\s{{M}}\right)+O(d^3) \ . \label{GM}
\end{align}
Here $d=d_g(z,w)$ is the geodesic distance between points with coordinates $z,w$, and $A_i$ are heat kernel coefficients. For constant scalar curvature they are given by
\begin{align}
A_0=1+\f{R}{24}d^2+\f{R^2}{640}d^4+O(R^3),\quad A_1=\f{R}{6}+\f{R^2}{120}d^2+O(R^3),\quad A_2=\f{R^2}{60}+O(R^3) \ .
\end{align}
Finally, the modified Bessel functions of the second kind $K_0, K_1, K_2$ can be defined by
\begin{align}
K_0(r)=\int_0^\infty \f{ds}{2s}e^{-s-\f{r^2}{4s}},\qquad K_1(r)=-K_0'(r),\qquad K_2(r)=2K_0''(r)-K_0(r) \ .
\end{align}
Function $K_0(r)$ has a logarithmic singularity at the origin ($\gamma_E\approx 0.577$ is Euler's constant) 
\begin{align}
K_0(r)=-\log r+\gamma_E-\log 2+ O(r)
\end{align}
and decays exponentially at $r\to \infty$ as is expected from the propagator of a massive field.

We will compute the loop integrals in the coordinate space. UV-divergences then appear as singularities of the propagator at coincident points. Following \cite{Zamolodchikov:2001ah} we regularize them by introducing the regularized propagator
\begin{align}
G^R_{{M}}(z)=\lim_{w\to z}\Big(G_{M}(z,w)+\log d(z,w)\Big) \ .
\end{align}
With this prescription all loop diagrams are finite. Explicitly the regularized propagator is
\begin{align}
G_{{M}}^R(z)=-\f12\log {M}+\log 2-\gamma_E+\f{R}{12{M}}+\f{R^2}{120{M}^2}+O(M^{-3}) \ . \label{GR}
\end{align}

\subsection{One loop}
We are now in a position to carry out the loop computations. The one-loop determinant $W$ is given by the partition function of the Gaussian field
\begin{align}
e^{W[g,N]}=\int \c{D}\chi\,\, \exp\left(-\f{1}{8\pi \be}\int_{\sph} \vol{g}{z}\left[ -\chi\D_g\chi +{M}\chi^2\right]\right) \ .
\end{align}
To compute it we note that
\begin{align}
\f{\p W[g,N]}{\p {M}}=-\f1{8\pi\be}\int_{\sph} \vol{g}{z} \Braket{\chi^2(z)}=-\f{1}{4\pi}\int_{\sph} \vol{g}{z} \,\,G^R_{{M}}(z)
\end{align}
and hence $W$ can be found by integrating the regularized propagator. The result is
\begin{multline}
W[g,N]=-\f{1}{4\pi}\int_{\sph} \vol{g}{z}\left[ -\f12{M} \log{M}+\f{1}{12}R\log{{M}}+\right.\\\left.\left(\f12+\log 2-\gamma_E\right){M}-\f1{120}\f{R^2}{{M}}\right]+O(N_H^{-2}) \label{WM}\ .
\end{multline}
Substituting \eqref{Mdef} and expanding in $N_H$ we find 
\begin{align}
W[g,N]\simeq \f{l_1(\be)}{2\pi N_H}\int_{\sph} \vol{g}{z}\,R^2+O(N_H^{-2})
\end{align}
with
\begin{align}
l_1(\be)=\f{(\be-2)(3\be-8)}{192}+\f{1}{480} \ .
\end{align}
\subsection{Two loops}
We now consider the two loop contributions. There are three connected diagrams at this order 
\begin{align}
&D^{(1)}_2=-\f1{4\pi}\f{\be}{6\pi}\int_{(\sph)^{2}} \s{g}d^2z\, \s{g}d^2w\,\,M^2 \left[G_M(z,w)\right]^3
&&\begin{tikzpicture}[baseline={([yshift=-.5ex]current bounding box.center)}, scale=0.25]
\draw (-6,0) -- (-6,0); % to align graph
\draw[thick] (-3,0) -- (3,0);
\draw[thick] (0,0) circle (3);
\end{tikzpicture}\label{D21}\\
&D^{(2)}_2=-\f1{4\pi}\f{\be}{4\pi}\int_{(\sph)^{2}} \s{g}d^2z\, \s{g}d^2w\,\, M^2 \left[ G^R_M(z)G_M(z,w)G^R_M(w)\right]
&&\begin{tikzpicture}[baseline={([yshift=-.5ex]current bounding box.center)}, scale=0.5]
\draw[thick] (-1,0) -- (1,0);
\draw[thick] (-2,0) circle (1);
\draw[thick] (2,0) circle (1);
\end{tikzpicture}\label{D22}\\
&D^{(3)}_2=\f1{4\pi}\f{\be}{2}\int_{\sph} \s{g}d^2z\,\, M\left[G^R_M(z)\right]^2
&&\begin{tikzpicture}[baseline={([yshift=-.5ex]current bounding box.center)}, scale=0.25]
\draw (-6,3) -- (-6,3); % to align graph
\draw[thick] (-2,0) circle (2);
\draw[thick] (2,0) circle (2);
\end{tikzpicture} \label{D23}
\end{align}
At large $M$ all integrals of this type localize near diagonals of the propagators and in the large $N_H$ expansion acquire a local form of eq. \eqref{Fexp}. In fact, one can formally solve equation \eqref{GMeq} in the large $M$ expansion as follows
\begin{align}
G_M(z,w)=\f{2\pi}{M-\D_g} \delta(z,w)=\f{2\pi}{M}\delta(z,w)+\f{2\pi}{M^2}\D_g\delta(z,w)+\dots \ . \label{Gformal}
\end{align}
This expansion in terms of delta-functions emphasizes the localization of the integrals. Technically \eqref{Gformal} gives a sufficient approximation if the propagator is to be integrated against smooth at $z=w$ functions. We can use it to show that diagrams \eqref{D22} and \eqref{D23} in fact cancel each other. Indeed, substituting \eqref{Gformal} in \eqref{D22} gives
\begin{align}
D_2^{(2)}=-\f1{4\pi}\f{\be}{2}\int_{\sph} \vol{g}{z}\,\,M\left[G^R_M(z)\right]^2-\f1{4\pi}\f{\be}{2}\int \s{g}d^2z\,\, G^R_M(z)\D_gG^R_M(z)+O(N_H^{-2})\ .
\end{align}
The first term here precisely cancels \eqref{D23} while all other involve derivatives of $G^R_M$ and vanish at constant curvature.

Thus the two-loop contribution comes entirely from \eqref{D21}. Formal expansion \eqref{Gformal} is not applicable to \eqref{D21} as it produces singularities of the type $\delta(0)$. We will take a more direct approach instead. Let us fix $z$ in \eqref{D21} and perform an integration over $w$. We assume that coordinate system $w$ is chosen so that the metric has the form
\begin{align}
2g_{w\bar w}=\f{1}{\left(1+\f{R}{8}r^2\right)^2}=1-\f{R}{4}r^2+\f{3R^2}{64}r^4+O(R^3) \ ,
\end{align}
where $r=|w-z|$. Expansion of the geodesic distance in these coordinates reads
\begin{align}
d(z,w)=r-\f{R}{24}r^3+\f{R^2}{320}r^5+O(R^3) \ .
\end{align}
Substituting this expansion into \eqref{GM} one finds \footnote{We note that as a non-trivial check for the correctness of these expansions one can verify that they satisfy equation $\p_M G_M(z,z')=-\f{1}{2\pi}\int \s{g}d^2w\,\, G_M(z,w)G_M(w,z')$ which can be derived by differentiating \eqref{GMeq} with respect to $M$.}
\begin{multline}
G_M(z,w)=K_0(\rho)+\f{R}{24 M}\left(\rho^2K_0(\rho)+\rho(2+\rho^2)K_1(\rho)\right)+\\\f{R^2}{5760 M^2}\left(\rho^2(24+9\rho^2+5\rho^4)K_0(\rho)-3\rho(\rho^4-8\rho^2-16)K_1(\rho)\right)+O(N_H^{-3}) \ , \label{GMexp}
\end{multline}
with $\rho=r\s{M}$. 

Finally, substituting \eqref{GMexp} into \eqref{D21} one finds
$D_{2}^{(1)}\simeq \f{\be l_2 }{2\pi N_H} \int \s{g}d^2z R^2+O(N_H^{-2})$ 
with 
\begin{multline}
l_2= -\f{1}{23040}\int_0^\infty dr\, r^2 K_0(r) \left[10 r \left(r^2+2\right)^2 K_1(r){}^2+\right.\\\left.r \left(5 r^4+49 r^2+24\right) K_0(r){}^2+\left(-43 r^4-56 r^2+48\right) K_1(r)K_0(r)\right] \label{I}\ .
\end{multline}
Numerically evaluating \eqref{I} gives
\begin{align}
l_2\approx -1.688139\times 10^{-3} \ . \label{l2}
\end{align}
\subsection{Three loops}
There are fifteen connected three-loop diagrams. However, all diagrams involving regularized propagators (i.e. self-loops) cancel up to order $O(N_H^{-2})$ by the same mechanism that canceled \eqref{D22} and \eqref{D23}. The following four diagrams remain
\begin{align}
&D_3^{(1)}=\f{\be }{48\pi^2} \int M^2 \left[G_{12}^4\right] &&\begin{tikzpicture}[baseline={([yshift=-.5ex]current bounding box.center)}, scale=0.25]
\draw (-6,0) -- (-6,0); % to align graph
\draw[thick] (0,0) ellipse (3 and 1.2);
\draw[thick] (0,0) circle (3);
\end{tikzpicture}\label{D31}\\
&D_3^{(2)}=-\f{\be^2 }{16\pi^3} \int M^3 \left[G_{13}^2G_{13}^2G_{23}\right] &&\begin{tikzpicture}[baseline={([yshift=-.5ex]current bounding box.center)}, scale=0.25]
\draw (-6,0) -- (-6,0); % to align graph
\draw[thick] (0,0) circle (3);
\draw[thick] (-3,0) .. controls (-1,0) and (0,1) .. (0,3);
\draw[thick] (3,0) .. controls (1,0) and (0,1) .. (0,3);
\end{tikzpicture}\label{D32}\\
&D_3^{(3)}=\f{\be^2 }{64\pi^4} \int M^4 \left[G_{12}G_{13}^2G_{24}^2 G_{34}\right] &&\begin{tikzpicture}[baseline={([yshift=-.5ex]current bounding box.center)}, scale=0.25]
\draw (-6,0) -- (-6,0); % to align graph
\draw[thick] (0,0) circle (3);
\draw[thick] (-2.12,2.12) .. controls (0,1) .. (2.12,2.12);
\draw[thick] (-2.12,-2.12) .. controls (0,-1) .. (2.12,-2.12);
\end{tikzpicture}\label{D33}\\
&D_3^{(4)}=\f{\be^2 }{96\pi^4} \int M^4 \left[G_{12}G_{13}G_{14}G_{23}G_{24}G_{34}\right]
&&\begin{tikzpicture}[baseline={([yshift=-.5ex]current bounding box.center)}, scale=0.25]
\draw (-6,0) -- (-6,0); % to align graph
\draw[thick] (0,0) circle (3);
\draw[thick] (-3,0) -- (3,0);
\draw[thick] (0,-3) -- (0,-0.5);
\draw[thick] (0, 3) -- (0,0.5);
\end{tikzpicture}\label{D34}
\end{align}
where for notational clarity we abbreviated $G_{i,j}=G_M(z_i,z_j)$ and  omitted the integration measures. Introduce also coefficients $l_3^{(k)}$ by
\begin{align}
D_3^{(k)}\simeq \f{\be^2 l_3^{(k)}}{2\pi N_H}\int \s{g}d^2z\,\, R^2 +O(N_H^{-2}) .
\end{align}
Evaluation of the diagram \eqref{D31} reduces to a one-dimensional integral similar to \eqref{D21} and gives numerically
\begin{align}
l_3^{(1)}\approx 1.45\times10^{-3} \ .
\end{align}
In contrast, other diagrams result in multidimensional integrals which are harder to treat numerically with high precision. With methods available to us we find
\begin{align}
l_3^{(2)}\approx -4.82\times10^{-3} ,\qquad l_3^{(3)}\approx 2.16\times10^{-3} ,\qquad l_3^{(4)}\approx 0.93\times10^{-3}\ ,
\end{align}
with accuracy of each number about $1\%$. Combining all contributions together gives
\begin{align}
l_3\approx  \lthree \ . \label{l3}
\end{align}

Putting all loop contributions together we find the following result for $c_1(\be)$
\begin{align}
c_1(\be)=-\frac{1}{8 \beta }+\frac{101}{480}+\beta  \left(l_2-\frac{5}{48}\right)+\beta ^2 \left(l_3+\frac{1}{64}\right)+O\left(\be^3\right)  \label{c1exp}\ .
\end{align}
Rational coefficients in this expression come from the saddle-point and the one-loop determinant and the coefficients $l_2, l_3$ arise from the calculation of higher loop diagrams.
 
If we compare this numerical result to the known $\be=1$ value Eq.\ \eqref{c1} we find an agreement with $98\%$ accuracy. 

\section{Summary \label{sec disc}}
We have studied the generating functional for the Laughlin state which is given by a Coulomb-type integral \eqref{Zdef}. The result is most economically written in terms of the effective magnetic field $H=N_h h$ introduced in \eqref{Hdef}
\begin{multline}
\log Z[W]\equiv F[g_H]=\\-\f{N_H^2}{2\pi\be}\iint_{(\sph)^2} \D_h^{-1}-\f{N_H}{2\pi}\f{\be-2}{2\be}\iint_{(\sph)^2} \D_h^{-1}R_h+\f{c}{96\pi}\iint_{(\sph)^2} R_h\D_h^{-1}R_h+\\\f{c_1(\be)}{2\pi N_H}\int_{\sph}\vol{g_h}{z} R_h^2+O(N_H^{-2}) \label{maindisc} \ .
\end{multline}
Recall that $g_h=h\,g, R_h=h^{-1}\left(R-\D_g \log h\right)$ and $c$ is given by
\begin{align}
c=1-3\f{(\be-2)^2}{\be} \ . \label{cdisc}
\end{align}
When expressed in terms of the magnetic field $B$, scalar curvature $R$ of the original Riemannian metric $g$ and the corresponding Laplace-Beltrami operator $\Delta_g$, the result \eqref{Hdef} for the generating functional reads

\hspace*{-1.5cm}\vbox{
\begin{multline}
\log Z=-\f1{2\pi}\sigma_H\iint_{(\sph)^2} B\D_g^{-1}B-\f1{2\pi}2\varsigma_H\iint_{(\sph)^2} B\D_g^{-1}R+\frac1{2\pi}\cdot\f{c_H}{48}\iint_{(\sph)^2} R\D_g^{-1}R\\
-\f{1}{2\pi}\int_{\sph} \s{g}d^2z\left[\frac{2-\beta}{2\beta} B\log B+\left(\f1{24}-\frac{(\beta-2)(\beta-2s)}{8\beta}\right)R\log B+\frac c{48}(\log B)\D_g\log B\right]\\+\f{1}{2\pi}\int_{\sph}\vol{g}{z}\big(c_1(\beta)(\D_g\log B)^2B^{-1}+c_1'(\beta)R(\D_g \log B)B^{-1}+c_1''(\beta)R^2B^{-1}\big)+O(N^{-2})
\label{main1}
\end{multline}
}

\noindent Here the coefficients $c_1',c_1''$ are related to $c_1$ as
\begin{align}
&c_1'(\be)=2c_1(\be)+c\f{1-s}{48}\label{c1exp'}\\
&c_1''(\be)=c_1(\beta)+\f{(1-s)\Big(3\be^2-(10+3s)\be+6(s+1)\Big)}{48\be}
\label{c1exp''}
\end{align}
and the Laurent expansion of the coefficient $c_1(\be)$ near $\beta=0$ is given in Eq.\ \eqref{c1exp}.

\subsection*{Acknowledgments}	
We thank Sylvain Ribault for useful discussions.
The work of NN is partly funded by DFG projects CRC/TRR 191 and SFB/TRR 183. The work of SK has benefitted from support provided by the University of Strasbourg Institute for Advanced Study (USIAS) for a Fellowship, within the French national programme “Investment for the future” (IdEx-Unistra), by the University of Strasbourg IdEx program and by the RFBR grant 18-01-00926.

\end{document}